  \documentclass[final,1p,times,authoryear]{elsarticle}

\usepackage{amssymb}
\usepackage{amsmath}
\usepackage{bm}
\usepackage{caption}
\usepackage{subcaption}
\usepackage{xcolor}
\usepackage[table]{xcolor}
\usepackage{longtable}

\newcommand{\qt}[1]{\textcolor{black}{#1}}

\journal{International Journal of Heat and Fluid Flow}

\begin{document}

\begin{frontmatter}



\title{The effect of bubble induced turbulent structures on the mass transfer of non-spherical bubbles}


\author[second]{V.J.G. Dijke}
\author[second]{R.G. Meijer}
\author[second,first]{M.W. Baltussen\corref{cor1}}
\affiliation[second]{organization={Department of Chemical Engineering and Chemistry, Multiphase Flows for Energy Applications, Eindhoven University of Technology}, address={P.O. Box 513}, poscode={5600 MB}, city={Eindhoven}, country={The Netherlands}}
\affiliation[first]{organization={Eindhoven Institute for Renewable Energy Systems, Eindhoven University of Technology}, city={Eindhoven},address={P.O. Box 513}, postcode={5600  MB}, country={the Netherlands}}
\ead{m.w.baltussen@tue.nl}
\cortext[cor1]{Corresponding author}

\begin{abstract}
Although mass transfer from bubbles to liquid is essential for the prediction of the efficiency of reactors, the mass transfer from bubbles is not fully understood. To determine the effect of the local velocity profile on the mass transfer for a wobbling bubble with an E\"otv\"os number of 2 and a Morton number of $10^{-11}$, 15 simulations were performed with a Front-Tracking method using a sub-grid scale model for the mass transfer in the vicinity of the interface. The vortical structures created by the bubble are influenced by the exact physical properties chosen for the liquid and gas. These changes in the vortical structures also resulted in changes in mass transfer. In addition, the vortical structures created transport barriers between the wake and the bulk of liquid, which were identified by the high-value Finite Time Lyapunov Exponents. These barriers prevent convective mass transfer from the bubble wake to the bulk of the liquid. Therefore, mass transfer from the gas phase to the bulk liquid should take into account both the mass transfer from the gas to the liquid and the transfer from the wake to the bulk of the liquid.
\end{abstract}


\end{frontmatter}

\section{Introduction}
In industrial applications, bubbles are essential to either mix the liquid and/or contact the gaseous and liquid components. The liquid is mixed by the rise of the bubbles, which create turbulent structures in the wake when the bubbles are large enough (\textit{i.e.} if the bubbles are large enough to have helical or wobbling pathways) or by the interactions of the bubbles \citep{Radl2007FlowFluids, Timmermann2016InfluenceReaction}. TIn addition, the bubbles can provide gaseous components to the liquid (\textit{e.g.}, the supply of oxygen for bacteria in waste water treatment) or to remove gaseous products from the liquid phase (\textit{e.g.}, hydrogen bubble production in water electrolysis).

Although mass transfer from bubbles to the liquid is of utmost importance to the efficiency of these devices, mass transfer from bubbles is far from understood even for single bubbles \citep{Claassen2026MassEffects}. There is a consensus that the gas liquid mass transfer is governed by a combination of molecular diffusion and convection. When looking at the mass transfer from a bubble to the bulk of the liquid, several different zones of mass transfer can be identified: in the bulk of the liquid and gas, both molecular diffusion and convection are important, whereas in the thin film close to the interface, the mass transfer is governed only by the molecular diffusion. However, the precise interaction between these mechanisms is still unclear, while it is clear that the bubble behavior and its mass transfer are inherently correlated through the flow of the liquid around the bubble \citep{Yan2023MechanismsReview}.

To capture the mass transfer for non-spherical bubbles, a wealth of mass transfer models have been proposed by \textit{e.g.}, \citet{Lochiel_1964}, \citet{ Anderson1967FundamentalsVibrations}, \citet{Montes1999MassBioreactors}, \citet{Tsuchiya2003DynamicsSystem} and \citet{Brauer1971StoffaustauschChemischerReaktionen}. Note that these correlations determine the average mass transfer coefficients for these bubbles. \citet{Montes1999MassBioreactors}, \citet{Tsuchiya2003DynamicsSystem} and \citet{Bork2005TheSystems} acknowledge that the effect of capillary waves on the gas-liquid interface will change the mass transfer by effectively changing the local concentration profile \citep{Yan2023MechanismsReview}. However, the capillary waves on the surface also cause changes in the wake profile behind the bubble leading to turbulence like structures, also referred to as bubble induced turbulence. In bubble swarms, bubble-induced turbulent structures of the other bubbles will influence mass transfer from a bubble both by altering local concentration gradients and by modifying the bubble rise velocity \citep{Ma2025,Hessenkemper2025}.

Although the mass transfer of a single rising bubble in a quiescent fluid medium is not well understood, the effect of turbulent structures has been studied. These studies showed that the liquid flow field around the bubble significantly affects the mass transfer near the gas-liquid interface. The focus of these studies has been mainly on turbulent structures in the liquid and interfacial convection \citep{Zheng2024FlowTransfer}.  

In this work, we want to better understand the effect of the turbulent structures in the wake on the mass transfer and the concentration distribution around the bubble. To study the effect of the wake structure on mass transfer, we performed simulations with different physical properties for both gas and liquid, while keeping the dimensionless diameter of the bubble (indicated by the E\"otv\"os number, $Eo=\frac{g\Delta\rho d_{eq}^2}{\sigma}=2$, where $g$ is the gravitational acceleration in the main rise direction, $\Delta\rho$ the difference in the liquid and gas density, $d_{eq}$ the volume equivalent diameter and $\sigma$ the surface tension coefficient) and the dimensionless fluid properties (indicated by the Morton number, $Mo = \frac{g\mu_l^4\Delta\rho}{\rho_l^2\sigma^3}=10^{-11}$, where $\mu_l$ is the dynamic viscosity of the liquid and $\rho_l$ the density of the liquid) the same. This should result in only minor effects on the dimensionless bubble rise velocity, but could have significant effects on the velocity field in the vicinity of the gas-liquid interface \citep{Claassen2026MassEffects}. These local changes in the velocity field are expected to influence the wake pattern in this wobbling bubble and, therefore, allow us to determine the effect of the wake structure on the mass transfer without significant changes in the effective bubble rise velocity.

The paper will start with a short explanation of the methods used for determining the flow field and species transport around the bubble. Afterwards, the effect of the different vortical structures on the mass transfer from the bubble to the liquid will be discussed. This will be followed by a discussion on the effect of the wake structure on the concentration field around the bubble. Finally, the main conclusions of the work will be given.

\section{Methodology}
To compare different liquids and gasses while having the same E\"otv\"os and Morton numbers, this study will be performed using numerical simulations. In addition, numerical simulations allow for the quantification of the local mass transfer near the gas-liquid interface, which is not possible in experiments \citep{Yan2023MechanismsReview}. It should be noted that, using a similar methodology to that in the present work, the time averaged mass transfer into the mass transfer boundary layer can  also be determined from experiments assuming a stationary state of this mass transfer boundary layer \citep{HUANG2025105106}.

\subsection{Governing equations and numerical model}
All simulations in this work are performed by combining a Front Tracking method with a subgrid scale (SGS) model for the mass transfer in the vicinity of the bubble. A Front Tracking method is chosen because it directly tracks the gas-liquid interface, which is beneficial when determining the local mass transfer near the gas-liquid interface \citep{SintAnnaland2006}. The methods used in this work are based on the Front Tracking method developed by \citet{Dijkhuizen2010-model} and \citet{Roghair2011OnNumbers}.

\subsubsection{Fluid dynamics}
In this method, the incompressible continuity and Navier-Stokes equations are solved on the basis of a one-fluid approximation. 
\begin{equation}
    \nabla\cdot\mathbf{u} = 0
    \label{eq:continuity}
\end{equation}
\begin{equation}
    \rho \frac{\partial \mathbf{u}}{\partial t} = -\nabla p - \rho\nabla\cdot\left( \mathbf{u}\mathbf{u}\right) - \nabla\cdot\bm{\tau}+\rho \mathbf{g} + \mathbf{F}_{\sigma}
    \label{eq:NS}
\end{equation}
In this equation, $\mathbf{u}$ is the velocity, p is the pressure, $\bm{\tau}$ is the stress tensor, and $\mathbf{g}$ is the gravitational acceleration. The required fluid density $\rho$ and dynamic viscosity $\mu$ are determined based on the local volume fraction of the phase using normal and harmonic averaging, respectively. The local volume fraction is determined directly from the location of the interface, which is described by a Lagrangian mesh of triangular markers \citep{Dijkhuizen2010-model}. 

The last term in equation \ref{eq:NS}, $\mathbf{F}_\sigma$, is a force density to take into account the effect of surface tension and is determined directly from the triangular surface mesh. For all triangular elements, the force density can be determined by summing the tensile forces between all neighboring elements and the considered element. This tensile force between a marker and its neighbor depends on the joint tangent ($\mathbf{t}_{m,i}$) and the normal of the neighboring marker ($\mathbf{n}_{m}$) via \citep{Dijkhuizen2010-model}:
\begin{equation}
	\mathbf{F}_{\sigma, m} = \frac{1}{2}\sum_{i=1}^3\sigma (\mathbf{t}_{m,i} \times
\mathbf{n}_{m})
	\label{eq:Fsigma}
\end{equation}
Note that this tensile force method has been derived by \citet{TRYGGVASON2001708}, any interested reader is referred to this paper for the derivation of the method.

Equations \ref{eq:continuity} and \ref{eq:NS} are solved using a projection-correction algorithm. In this algorithm, an estimate of the velocity field is determined using equation \ref{eq:NS}. In this projection, all terms are treated explicitly except for the viscous term, which is treated semi-implicitly. The explicit part of the stress term is chosen such that it is small while ensuring that all components can be determined independently. For the time derivative, a first order Euler scheme is used, while the convective term and the viscous stress term are discretized using the Barton scheme and second order central differencing, respectively. The obtained estimate of the velocity field is corrected to ensure that the velocity field also satisfies equation \ref{eq:continuity}. The implicit part of the viscous stress and the correction step are solved using a block ICCG solver. Interested readers on the exact implementation of this pressure-correction algorithm are referred to \citet{Dijkhuizen2010-model}.

Based on the velocity field obtained, the interface mesh is advanced by individually moving all points of the triangulated mesh separately with a fourth order Runga-Kutta method. The velocity used for this advancement is determined by interpolating the velocity from the background Eulerian grid using cubic splines \cite{SintAnnaland2003, Dijkhuizen2010-model,Roghair_2016}. This separate movement of all points in the triangulated mesh causes the deformation of the gas-liquid interface, but also results in a deterioration of the mesh quality, which is improved by implementing a mass conservative re-meshing procedure \citep{Roghair_2016}. 

\subsubsection{Species transfer}
After the velocity field and the updated interface are obtained, the concentration profile can be determined using the following conservation equation for the concentration ($c$):

\begin{equation}
\frac{\partial c}{\partial {t}}+ (\mathbf{u}\cdot\nabla) c= D \nabla^2c
\label{eq:diffusion}
\end{equation}
where $D$ is the diffusion coefficient of the species in the continuous phase, which is considered constant. Note that there is a large difference between the concentration gradients in the bulk of the liquid (\textit{i.e.} away from the bubble) and near the gas-liquid interface. In the bulk of the liquid, the concentration gradients are sufficiently small to be captured by the hydrodynamics grid. In this region, equation \ref{eq:diffusion} is discretized using an implicit central differencing and an explicit van Leer scheme for the diffusion and the convective terms, respectively.

The concentration gradients in the vicinity of the interface are significantly higher than the concentration gradients in the bulk. To accurately represent these gradients in the vicinity of the interface, a subgrid scale (SGS) model acts within a distance of two grid cells in the normal distance from the interface. The SGS model solves a simplified conservation equation for each mesh element (\textit{i.e.} markers) assuming that the changes in the concentration only occur in the normal direction of the interface, the curvature does not influence the concentration profile and any tangential convection and diffusion are governed by the movement of the markers \citep{Aboulhasanzadeh2013ALiquids,Claassen_2019,Weiner2022AssessmentBubble}. The resulting conservation equation for each marker is the following:
\begin{equation}
    \frac{\partial c}{\partial t} = \mathbf{n} \gamma \frac{\partial c}{\partial \mathbf{n}} + D \frac{\partial^2c}{\partial \mathbf{n}^2},
    \label{eq:sgs_balance}
\end{equation}
In this equation, $\gamma$ is the strain rate experienced by the marker ($\gamma = - \frac{\partial \mathbf{u}_n}{\partial{\mathbf{n}}}$). BThis equation thus indicates that the concentration on each marker is changed by the compression or expansion of the boundary layer and diffusion in the normal direction.

Although the local concentration field in the SGS model is given by equation \ref{eq:sgs_balance}, it would introduce significant memory requirements to determine the concnetration profile for each marker. Therefore, the total mass per marker, $M_0$, is defined as the total amount of mass transferred in the SGS-region for this marker. Note that $M_0$ is a measure for the mass within a set distance from the gas-liquid interface, $\delta_0$ (which is two grid cells in this work). Based on the definition of $M_0$ and equation \ref{eq:sgs_balance}, a new conservation equation can be defined for each marker:
\begin{equation}
    \frac{dM_0}{dt}= -\gamma M_0 -D \left.\frac{\partial c}{\partial \mathbf{n}}\right|_0+\gamma c_{\delta_0}\delta_0 +D \left.\frac{\partial c}{\partial \mathbf{n}}\right|_{\delta_0}
    \label{eq:sgs_totalmassbalance}
\end{equation}

In the current work, the bubble is rising in an infinite clean liquid. Therefore, the concentration profile inside the boundary layer can be approximated by an error function:
\begin{equation}
    \frac{c(x)}{c_i} = \mathrm{erfc} \left( \sqrt{\pi} \frac{x}{\delta}\right)
    \label{eq:sgs_cprofile}
\end{equation}
Using this concentration profile in combination with equation \ref{eq:sgs_totalmassbalance}, the penetration depth $\delta$ can be determined with a Newton-Raphson solver. Based on $\delta$, the concentration at the edge of the SGS model, \textit{i.e.} at $\delta_0$, can be determined directly. The concentration at $\delta_0$ will serve as a boundary condition for the background concentration grid using an approach similar to the Immersed Boundary method of \citet{Deen2012}.

The mass on each marker element is changed not only by the mass transferred, but also by the changes in the marker elements. Therefore, $M_0$ needs to be updated when remeshing is performed. The exact effects of remeshing on $M_0$ depend on the performed remeshing operations. For details of redistribution of $M_0$ due to remeshing, the interested reader is referred to \citet{Claassen_2019}.

\subsection{Verification and validation}
The verification of the method has been performed in our previous work. The local Sherwood numbers are compared with the analytical solution  for a single bubble under Stokes flow and potential flow conditions ($Pe=ReSc=\frac{\rho_l v d_{eq}}{\mu_l}\frac{\mu_l}{\rho_l D}=1.6\cdot10^5$ with $Sc=\frac{\mu_l}{\rho_l D}=1.6\cdot10^8$ and $Sc=118$, respectively) \citep{Claassen_2019,Weiner2022AssessmentBubble}. In addition to comparison with these analytical solutions, the method was also validated with simulations with higher resolution for single spherical and slightly ellipsoidal bubbles \citep{Weiner2022AssessmentBubble}.

\subsection{Simulation settings}
In total, a set of 15 simulations have been performed. All simulations are performed with a domain with a uniform Cartesian grid with 200 grid cells in the gravitational direction (in this case the z-direction) and 150 grid cells in the other directions \citep{Dijkhuizen2010-drag}. The size of the grid is chosen to be 32 grid cells in the diameter of the equivalent sphere, which has been shown to be sufficient to capture both hydrodynamics and mass transfer for these bubbles in our previous work \citep{Claassen_2019,Weiner2022AssessmentBubble}. Note that the concentration profile away from the bubble can be significantly affected by the low resolution compared to the concentration gradients. This will lead to a thickening of the concentration profile due to numerical diffusion \citep{Weiner2022AssessmentBubble}. 

The boundary conditions for the domain are set to a zero gradient for the mass transfer and free slip for the velocity. The bubble is initialized as a sphere at 0.75 times the height of the computational domain in the gravitational direction and the center in the horizontal plane, which ensures a minimal effect of the boundary conditions on the bubble rise velocity and the transfer of mass \citep{Dijkhuizen2010-drag, Weiner2022AssessmentBubble}. Using a window shifting technique, the bubble is kept at the initial position, as all fields and the center of mass of the bubble are moved back to the original position when the center of mass of the bubble is moved by a single grid cell from its original position. Initially, the velocity and pressure are set to zero for the entire domain. The time step used in all simulations is between $5\cdot10^{-4}$s and $5\cdot10^{-6}$s depending on the exact settings. 

All simulations are run until the pseudo-stable state is reached. To determine the pseudo-stable state, all the maxima in $Re=\frac{\rho_lvd_{eq}}{\mu_l}$ (where $v$ is the bubble rise velocity) are determined, which are multiple as shown in figure \ref{fig:oscillations_Sh_Re}. When these maxima are within 2\% of the last maximum, the results are considered to be in a pseudo-stable state. Averaging occurs for the entire period starting and ending always at a maximum in $Re$.

Although all simulations are performed with the same dimensionless bubble size and dimensionless gas-liquid combination, the physical properties of the liquid are changed, resulting in different viscosity ($\kappa=\frac{\mu_g}{\mu_l}$) and density ($\lambda=\frac{\rho_g}{\rho_l}$) ratios. The resulting ranges in the physical properties are given in table \ref{tab:sumdimensionless_prop}. The range in $\kappa$ and $\lambda$ is chosen to include water-air ($\kappa=10^{-2}$ and $\lambda=10^{-3}$), the generally used ratios ($\kappa=10^{-2}$ and $\lambda=10^{-2}$) and liquid droplets in a liquid ($\kappa=10^{-1}$ and $\lambda=10^{-1}$). Note that in all simulations the density of the gas is considered to be 1000 kg m$^{-3}$. For detailed settings for all simulations, the reader is referred to table \ref{tab:allsim}. The Schmidt number used in the simulations is within the range of Schmidt numbers tested in the verification cases and is realistic value for a gas diffusing in a liquid. In addition, \citet{Weiner2022AssessmentBubble} and \citet{Claassen2026MassEffects} showed that the effects of changes in the Schmidt number are captured by a square-root dependence.

\begin{table}[]
    \centering
    \caption{Summary of the dimensionless physical properties used in the simulations. A detailed set of the physical properties per simulation is given in the data package.}
    \label{tab:sumdimensionless_prop}
    \begin{tabular}{lc}
         \hline
         Dimensionless property & Range\\
         \hline
         viscosity ratio, $\kappa$ & $\left[2\cdot10^{-2},5\cdot10^{-1}\right]$ \\
         density ratio, $\lambda$ & $\left[10^{-3},5\cdot10^{-1}\right]$ \\
         Eötvös number, $Eo=\frac{g\Delta \rho d_{eq}}{\sigma}$ & $2.0$ \\
         Morton number, $Mo = \frac{g \mu_l^4 \Delta \rho}{\rho_l^2\sigma^3}$ & $10^{-11}$ \\
         Schmidt number, $Sc$ & $10^3$ \\
         Equivalent diameter, $d_{eq}$ & $4$ mm\\
         Gravitational constant, $g$ & $(0,0,-10)$ m s$^{-2}$\\
         \hline
    \end{tabular}
\end{table}

\section{Results}
 This result section is divided into two sections: the effect of the vortex shedding on the Sherwood number and the effect of the vortical structures on the concentration profile in the wake of the bubble.

\subsection{Effect of vortex shedding}
The Sherwood number for this work is obtained using the following equation, where the mass transfer at the gas-liquid interface via the SGS model \citep{Bird2002TransportPhenomena, Aboulhasanzadeh2013ALiquids, Weiner2022AssessmentBubble}:
\begin{equation}
    Sh = \frac{d_{eq} k}{D}=\qt{\frac{\text{Total mass transfer rate}}{\text{Diffusion rate}}=\frac{d_{eq}}{D}\frac{D}{(c_0-c_\infty)}\frac{\int_S\left.\frac{dc}{dn}\right|_0dS}{S}=}\frac{d_{eq}}{c_0-c_\infty}\frac{\sum_{m\in M}\left.\frac{dc}{dn}\right|_{0,m}A_m}{\sum_{m\in M}A_m}
    \label{eq:Sh_def}
\end{equation}
where $S$ is the area of the gas-liquid interface, $A_m$ the actual area of each marker $m$ and $M$ are all markers on the interface. Note that $c_\infty$, the far-field concentration, is zero for all cases considered.

\begin{figure}
    \centering
    \includegraphics[width=0.47\linewidth]{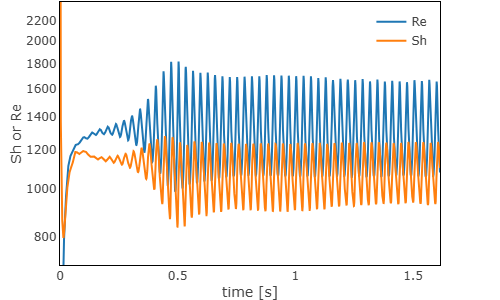}
    \includegraphics[width=0.47\linewidth]{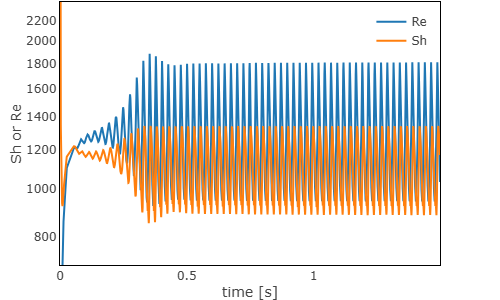}

    \caption{The time dependent behavior of the Reynolds and the Sherwood number for case 5 ($\kappa=\frac{\mu_g}{\mu_l}=10^{-1}$ and $\lambda=\frac{\rho_g}{\rho_l}=2\cdot10^{-1}$) and case 12 ($\kappa=\frac{\mu_g}{\mu_l}=10^{-1}$ and $\lambda=\frac{\rho_g}{\rho_l}=2\cdot10^{-3}$).}
    \label{fig:oscillations_Sh_Re}
\end{figure}

In figure \ref{fig:oscillations_Sh_Re}, the Reynolds number shows that in the initial period the bubbles have a steady acceleration and then transition to a pseudo-steady turbulent regime \citep{Jin2019DirectFlows}. The oscillations of the bubble in the turbulent regime are regular and have a clearly defined frequency and amplitude. Comparing the two cases, which vary by the density of the gas and as $Mo$ and $Eo$ are constant also the surface tension coefficient, the effective rise velocity is similar (\textit{i.e.}, $Re_{\text{case 5}}=1292$ and $Re_{\text{case 12}}=1254$), but the amplitude and frequency of the oscillations are affected by the change in the density of the gas and the surface tnesion coefficient.

When considering all cases listed in table \ref{tab:allsim}, it can be concluded that changes in physical properties have only a minor effect on the terminal rise velocity of the bubble (\textit{i.e.}, the change in the Reynolds number is less than 3\%). However, the frequency of the oscillations increases with a decrease in the dynamic viscosity of the gas phase, while there is only a small effect of the density of the gas phase. The amplitude of the oscillations seems to correlate with the exact Reynolds number and the used surface tension coefficient (\textit{i.e.} the deformability of the surface). This indicates that, although the average Reynolds number is unaffected, the wake structure is significantly influenced by changes in the physical properties of the gas and the liquid. 

In addition to the temporal variation in Reynolds number, figure \ref{fig:oscillations_Sh_Re} also shows the temporal variations of the Sherwood number. Comparing the results of the Sherwood number and the Reynolds number, it can be concluded that the oscillations of the Sherwood number are the same as the oscillations of the Reynolds number, which is confirmed for all simulations by the same frequencies observed for both dimensionless numbers in table \ref{tab:allsim}. However, there is a shift in the oscillations, which can be explained by the highest bubble rise velocities occurring when the bubble is in semi-stationary rise conditions, while the bubble rise velocity is decreased when the bubble turns. However, at these turning points, the highest surface renewal rates are obtained, and thus the highest mass transfer is expected. It confirms the theory of \citet{Tsuchiya2003DynamicsSystem}, \citet{Bork2005TheSystems} and \citet{Jin2023EffectsStudy} that the oscillations in the bubble rise path influence the local Sherwood number. Therefore, it can be concluded that the oscillation frequency for the Sherwood number also shows a dependence on the dynamic viscosity ratio. 

When comparing the average Sherwood number, only minor changes are observed for the different cases (approximately 8\% difference between all cases). These changes can partly be attributed to the changes in the bubble rise velocity but seem to be also correlated with the changes in the surface tension coefficient. A similar dependence is obtained for the variation of the amplitude of the oscillations in the Sherwood number. 

Although the number of simulations is insufficient to provide closure relations, the analysis above shows that the effects of the physical properties on the bubble rise velocity are minor, as also indicated by the Grace diagram \citep{Grace1973, Grace_1976}. However, the oscillations of the Reynolds number are affected by the physical properties of the gas and the liquid. The Sherwood number is affected by changes in the Reynolds number and physical properties both for the average value and the oscillations.

\subsection{Effect of the wake profile}
To determine the effect of the wake profile on the mass transfer to the bulk of the liquid, we studied the coherent structures in the wake of the bubble. To quantify these structures in the wake, the Finite-Time Lyapunov Exponent (FTLE) fields are determined, which provide a measure for the local divergence and convergence of the flow field based on the local velocity profile. The FTLE field is a Lagrangian analysis of the flow field, which provides the areas of the most attracting and repelling regions in the flow field by the highest values in the backward and forward FTLE field, respectively \citep{Shadden2005DefinitionFlows,Farazmand2012ComputingTheory,Haller2015LagrangianStructures,Kameke2019HowExample,Kursula2022UnsteadyReactor,Lagares2023ALayers}.

To determine the FTLE fields,  150 x 150 x 200 particles are carried by the trilinear interpolated velocity based on the simulated velocity field forward or backward in time by $\tau$. This advection is based on a simple Euler discretization, as \citet{Lagares2023ALayers} showed that the results are not sensitive to the discretization scheme used. Based on the movement of the particle and its neighbors, the Jacobian of the flow map, $\mathbf{J}$, is calculated as a 3 by 3 matrix:
\begin{equation}
    J_{i,j} = \frac{x_i(\mathbf{x}_0+\varepsilon\mathbf{e}_j)-x_i(\mathbf{x}_0-\varepsilon\mathbf{e}_j)}{2\varepsilon}
\end{equation}
In this equation, $x_i$ is the final position of the particle in one direction, $\mathbf{x}_0$ is the initial position of the same particle, $\mathbf{e}$ is the unit vector and $\varepsilon$ is a small offset. Based on this Jacobian, the Right Cauchy-Green Deformation tensor, $\Delta$, can be determined as:
\begin{equation}
    \Delta = \mathbf{J}^T\mathbf{J}
\end{equation}
The largest eigenvalue of the Right Cauchy-Green Deformation tensor determines the maximum squared stretching of the considered fluid element. With this largest eigenvalue $\lambda_{max}$, the FTLE component can be determined using the following:
\begin{equation}
    \text{Forward FTLE: } \Lambda^+(\mathbf{x}_0, \tau) = \frac{\ln(\lambda_{max}(\Delta))}{|\tau|}
\end{equation}
\begin{equation}
    \text{Backward FTLE: } \Lambda^-(\mathbf{x}_0, -\tau) = \frac{\ln(\lambda_{max}(\Delta))}{|\tau|}
\end{equation}
In this work, the FTLEs are calculated from a time instance of the velocity field, \textit{i.e.}, the time dependence of the velocity field is not taken into account in the calculation of the FTLE \citep{Lagares2023ALayers}.
    
\begin{figure}[]
    \centering
    \begin{subfigure}[]{0.45\textwidth}
        \centering
        \includegraphics[width=\textwidth]{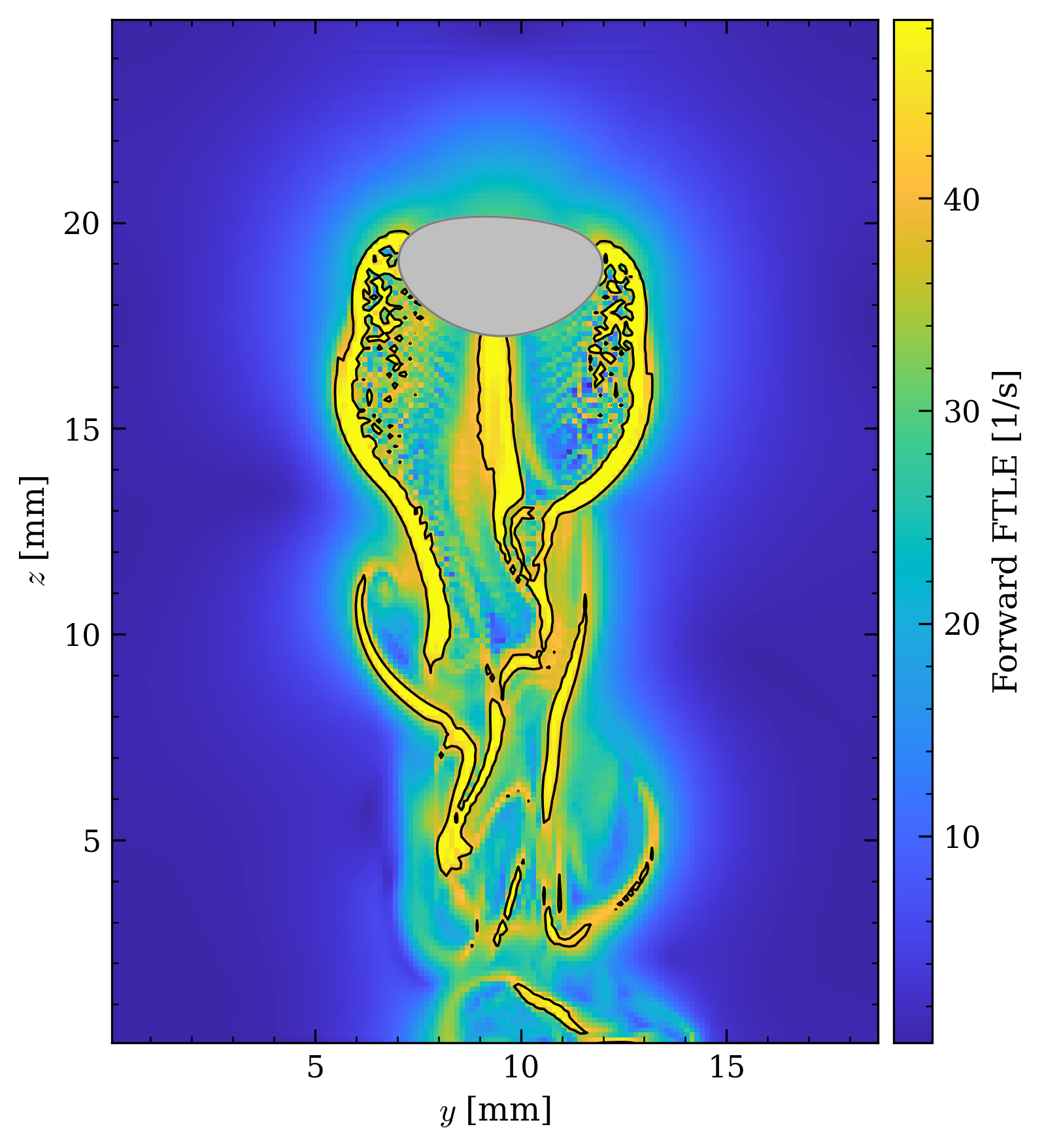}
        \caption{}
    \end{subfigure}
    \begin{subfigure}[]{0.45\linewidth}
        \centering
        \includegraphics[width= \textwidth]{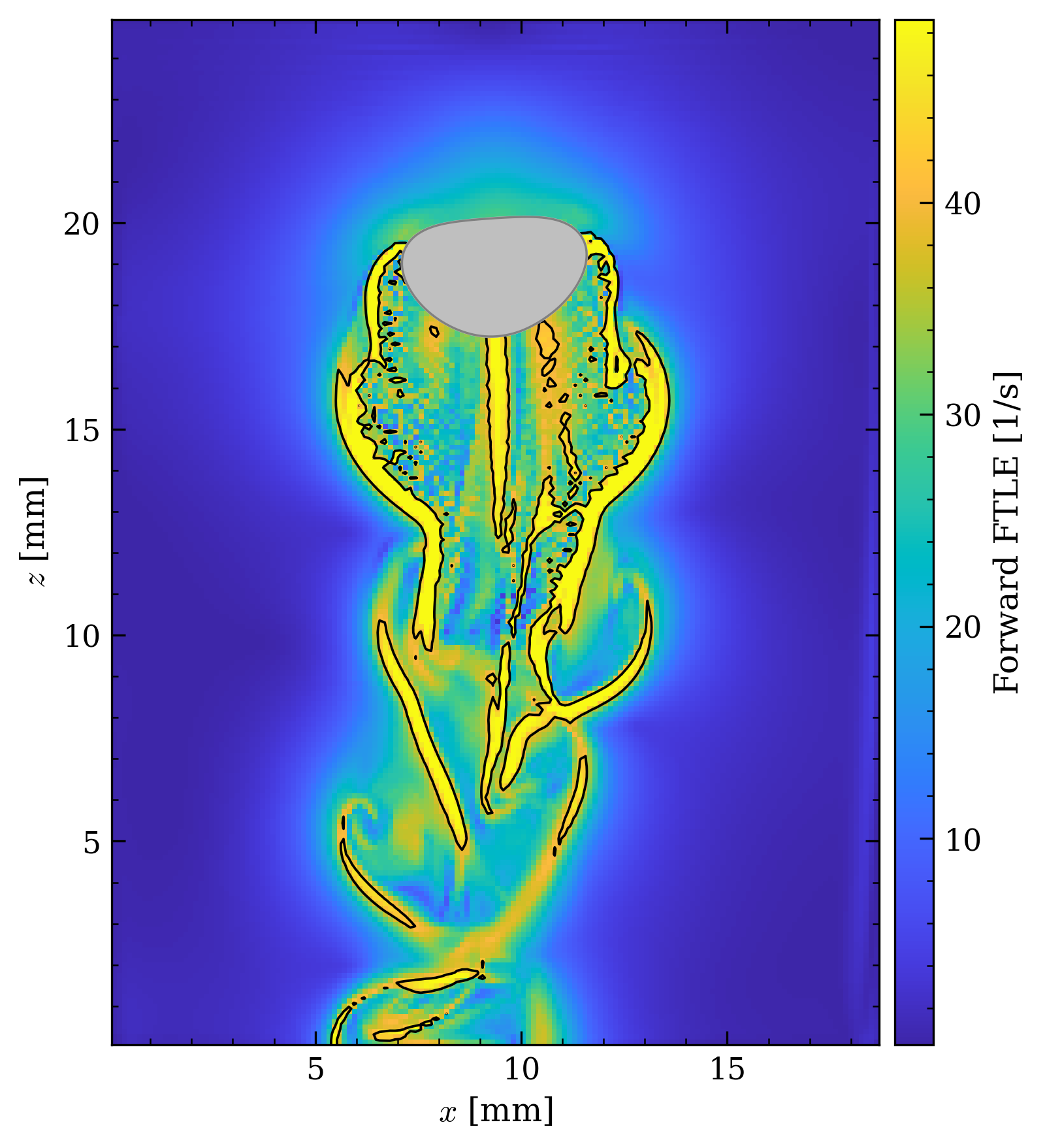}
        \caption{}
    \end{subfigure}
        \begin{subfigure}[]{0.45\textwidth}
        \centering
        \includegraphics[width=\textwidth]{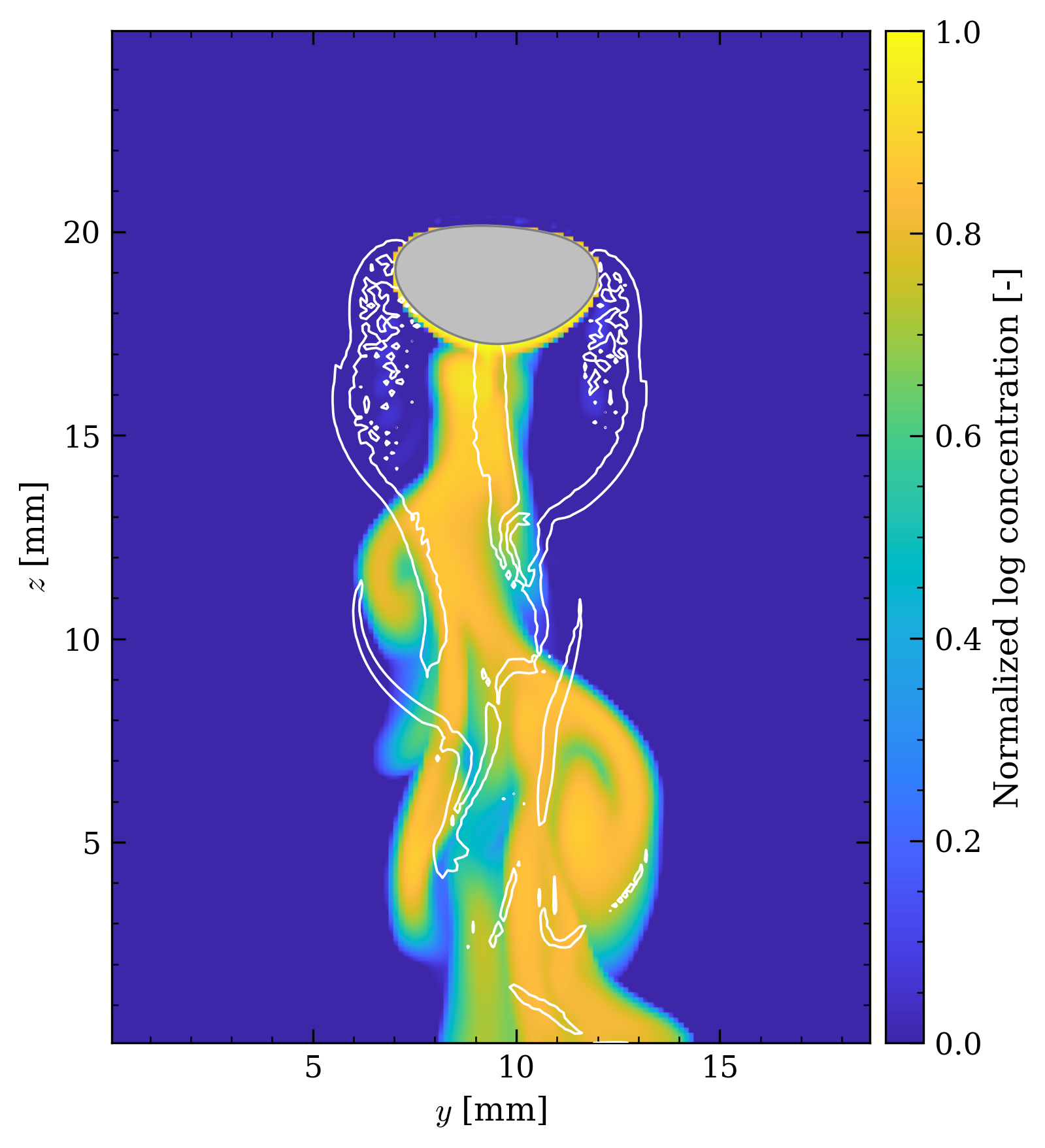}
        \caption{}
    \end{subfigure}
    \begin{subfigure}[]{0.45\linewidth}
        \centering
        \includegraphics[width= \textwidth]{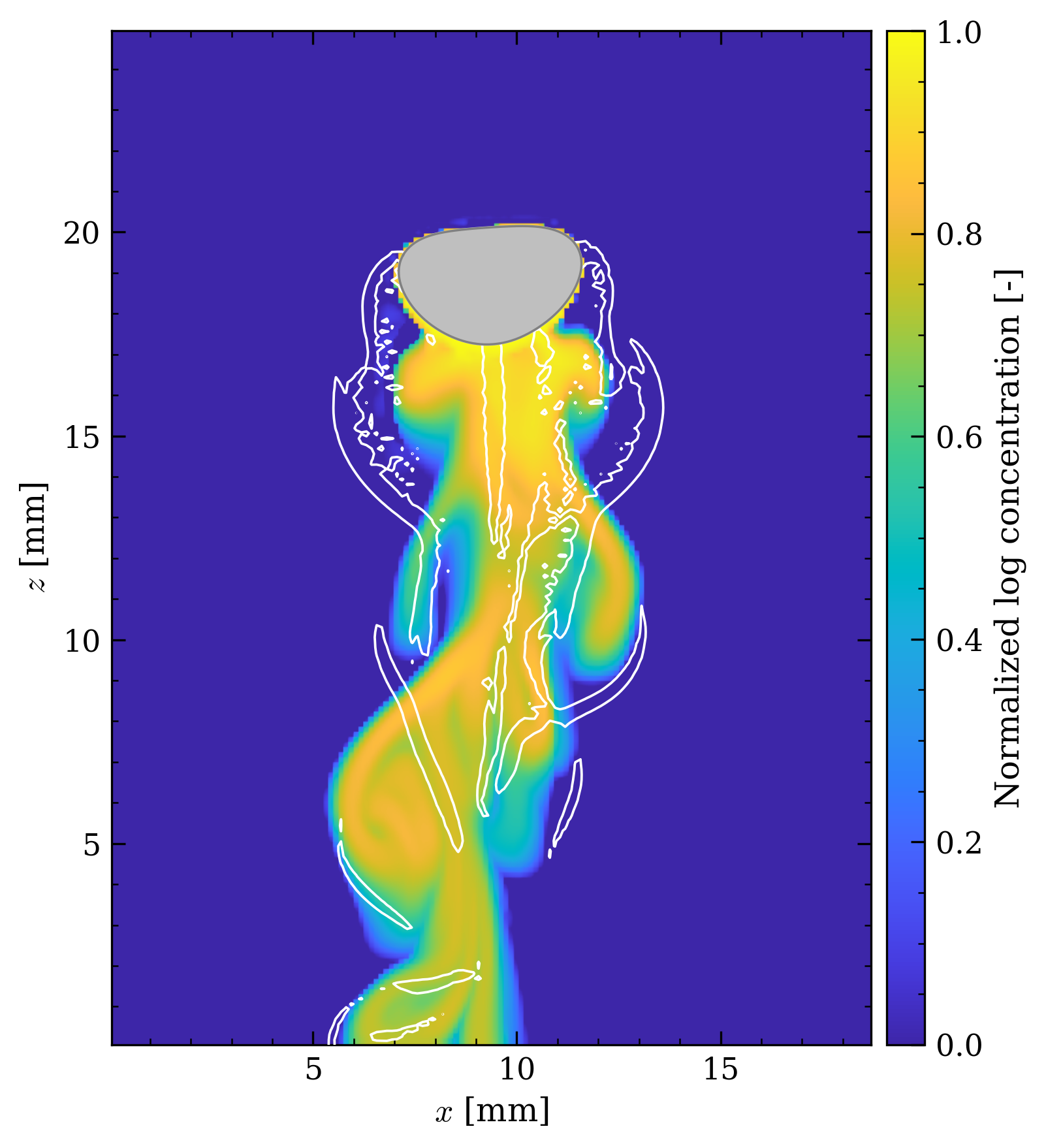}
        \caption{}
    \end{subfigure}
    \caption {The forward Finite-Time Lyapunov Exponent (FTLE) field (top row) and the concentration profiles (bottom row) for case 5 ($\kappa=\frac{\mu_g}{\mu_l}=10^{-1}$ and $\lambda=\frac{\rho_g}{\rho_l}=2\cdot10^{-1}$). Both columns present one direction, which are in perpendicular to each other. The lines in all figures indicate the FTLE ridges evaluated as the 5\% highest FTLE values.}
    \label{fig:case5_visualisation_fwd}
\end{figure}

\begin{figure}[]
    \centering
    \begin{subfigure}[]{0.45\textwidth}
        \centering
        \includegraphics[width=\textwidth]{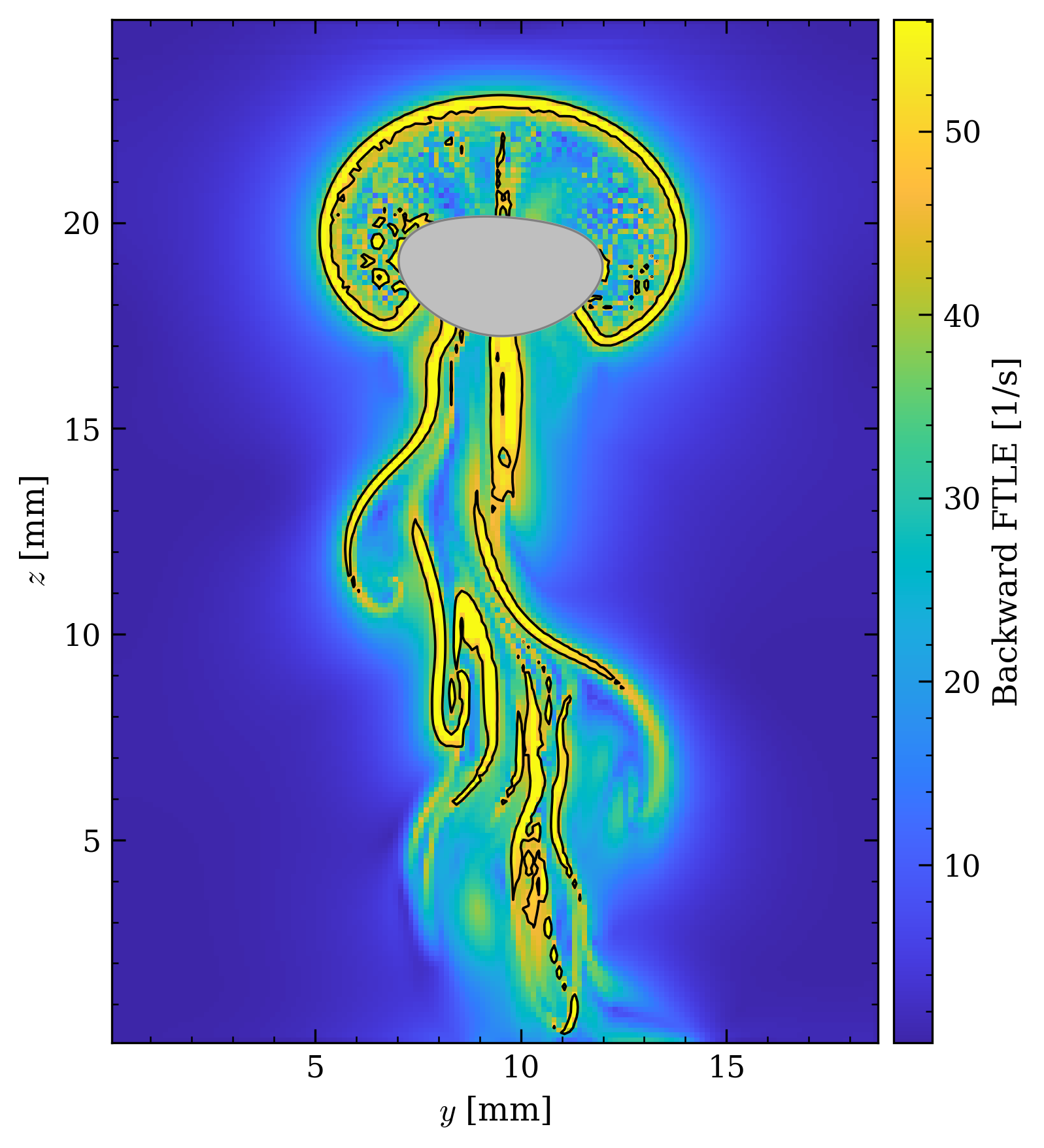}
        \caption{}
    \end{subfigure}
    \begin{subfigure}[]{0.45\linewidth}
        \centering
        \includegraphics[width= \textwidth]{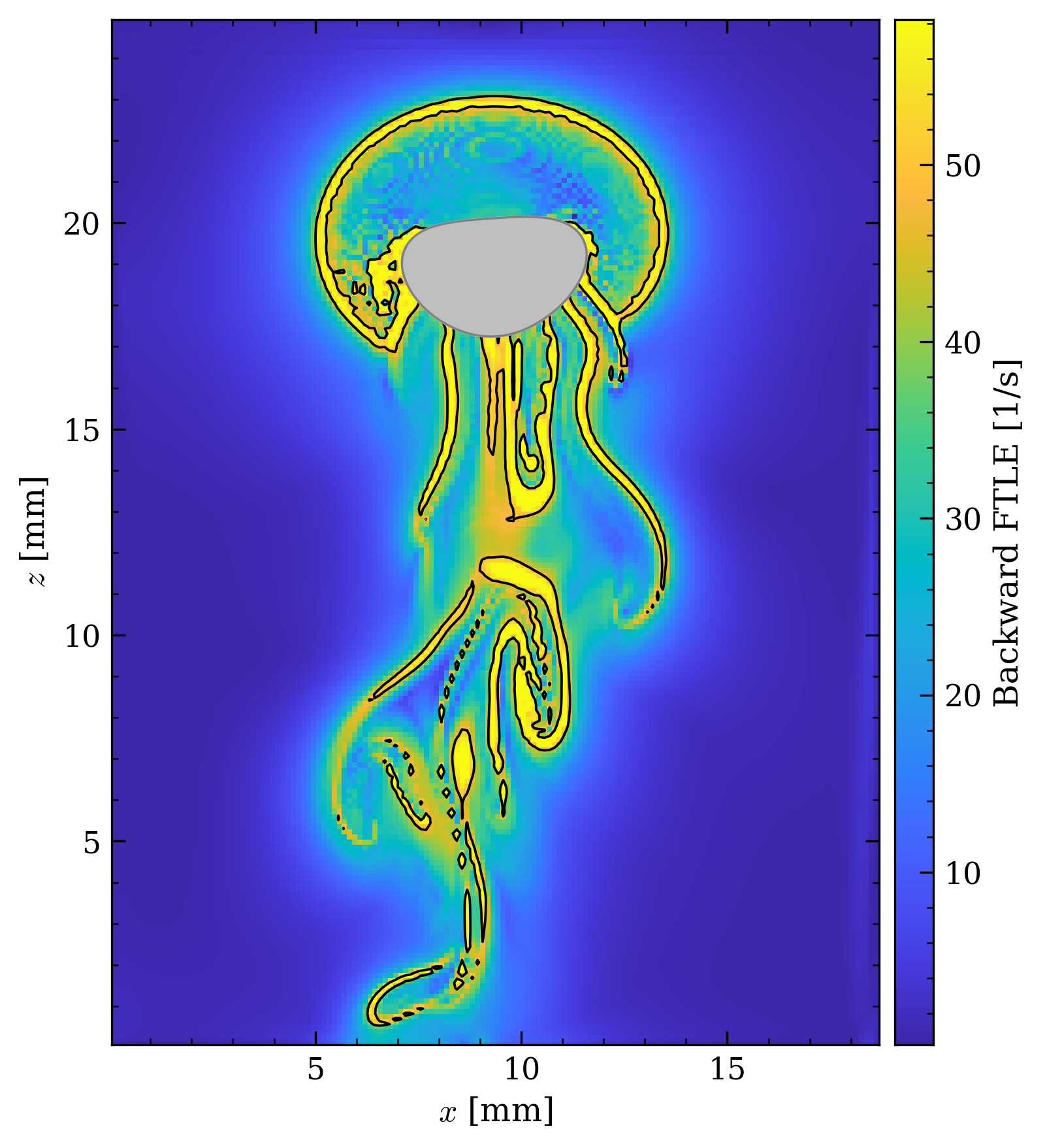}
        \caption{}
    \end{subfigure}
        \begin{subfigure}[]{0.45\textwidth}
        \centering
        \includegraphics[width=\textwidth]{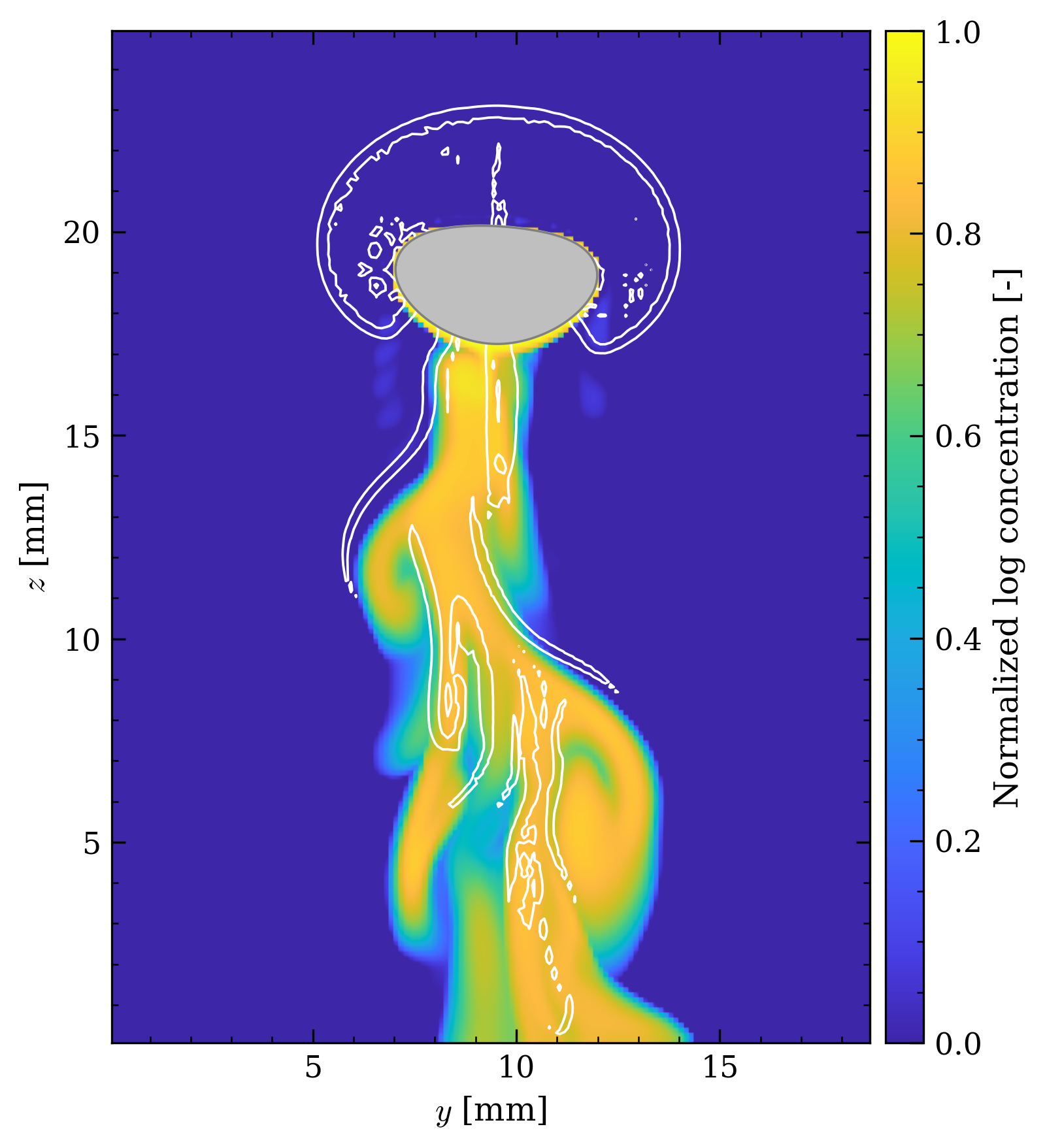}
        \caption{}
    \end{subfigure}
    \begin{subfigure}[]{0.45\linewidth}
        \centering
        \includegraphics[width= \textwidth]{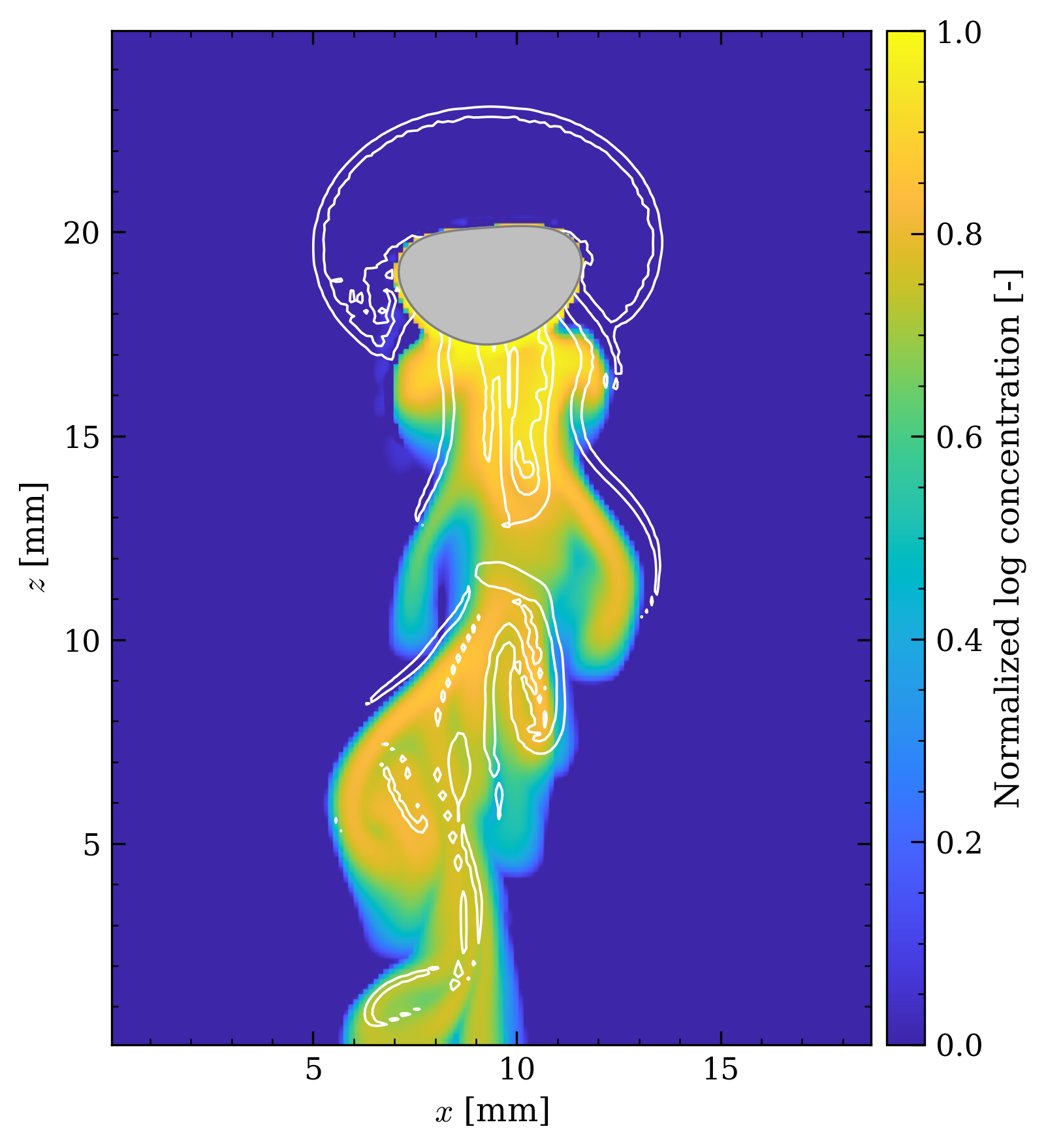}
        \caption{}
    \end{subfigure}
    \caption{The backward Finite-Time Lyapunov Exponent (FTLE) field (top row) and the concentration profiles (bottom row) for case 5 ($\kappa=\frac{\mu_g}{\mu_l}=10^{-1}$ and $\lambda=\frac{\rho_g}{\rho_l}=2\cdot10^{-1}$). Both columns present one direction, which are in perpendicular to each other. The lines in all figures indicate the FTLE ridges evaluated as the 5\% highest FTLE values.}
    \label{fig:case5_visualisation_bwd}
\end{figure}

First, the forward and backward FTLE fields in two perpendicular planes are shown for one instance for case 5 in the top row of figures \ref{fig:case5_visualisation_fwd} and \ref{fig:case5_visualisation_bwd}, respectively. It should be noted that a high value of the forward FTLE is considered to be a high stretching region, \textit{i.e.}, the fluid particles in these regions can be considered repelling, while high values of the backward FTLE are associated with attracting particles. Although the nature of high values in the FTLE is different for the backward and forward FTLE fields, both are considered barriers to transport \citep{Kursula2022UnsteadyReactor,Kameke2019HowExample}.

In both the forward and backward FTLE, predominantly vertical structures are obtained, which indicates that the horizontal mixing of the liquid is prevented by the vortical structures in the field. Note that the structures are influenced by the boundary conditions at the bottom of the simulation domain, where the alignment in the vertical direction is also lost. In addition, clear regions enclosed by high FTLE values are observed in both forward and backward FTLE fields, indicating that turbulent structures are isolating parts of the liquid volume preventing mixing with the environment. In other words, turbulent structures are created that are coherent in the flow field. 

In the bottom rows of figures \ref{fig:case5_visualisation_fwd} and \ref{fig:case5_visualisation_bwd}, the concentration field in the same direction and at the same instance are plotted with the FTLE ridges, which are created at the 5\% highest FTLE values (these are also visualized in the FTLE fields for comparison reasons). From figure \ref{fig:case5_visualisation_bwd}, it can be concluded that the concentration field is well described by the backward FTLE ridges. Similar to the experiments of \citet{Kursula2022UnsteadyReactor} and \citet{Kameke2019HowExample}, the high concentration regions are bounded by the FTLE ridges, which is logical as the backward FTLE ridges are an indication of the transport barriers in backward time and thus indicate how the flow has ordered in the past \citep{Kameke2019HowExample}. In figure \ref{fig:case5_visualisation_fwd}, the FTLE ridges in the forward FTLE indicate that the turbulent structures in the wake of the bubble will separate parts of the wake from their core, which also decreases the mixing within the wake. 


\begin{figure}[]
    \centering
    \begin{subfigure}[]{0.32\linewidth}
        \centering
        \includegraphics[width= \textwidth]{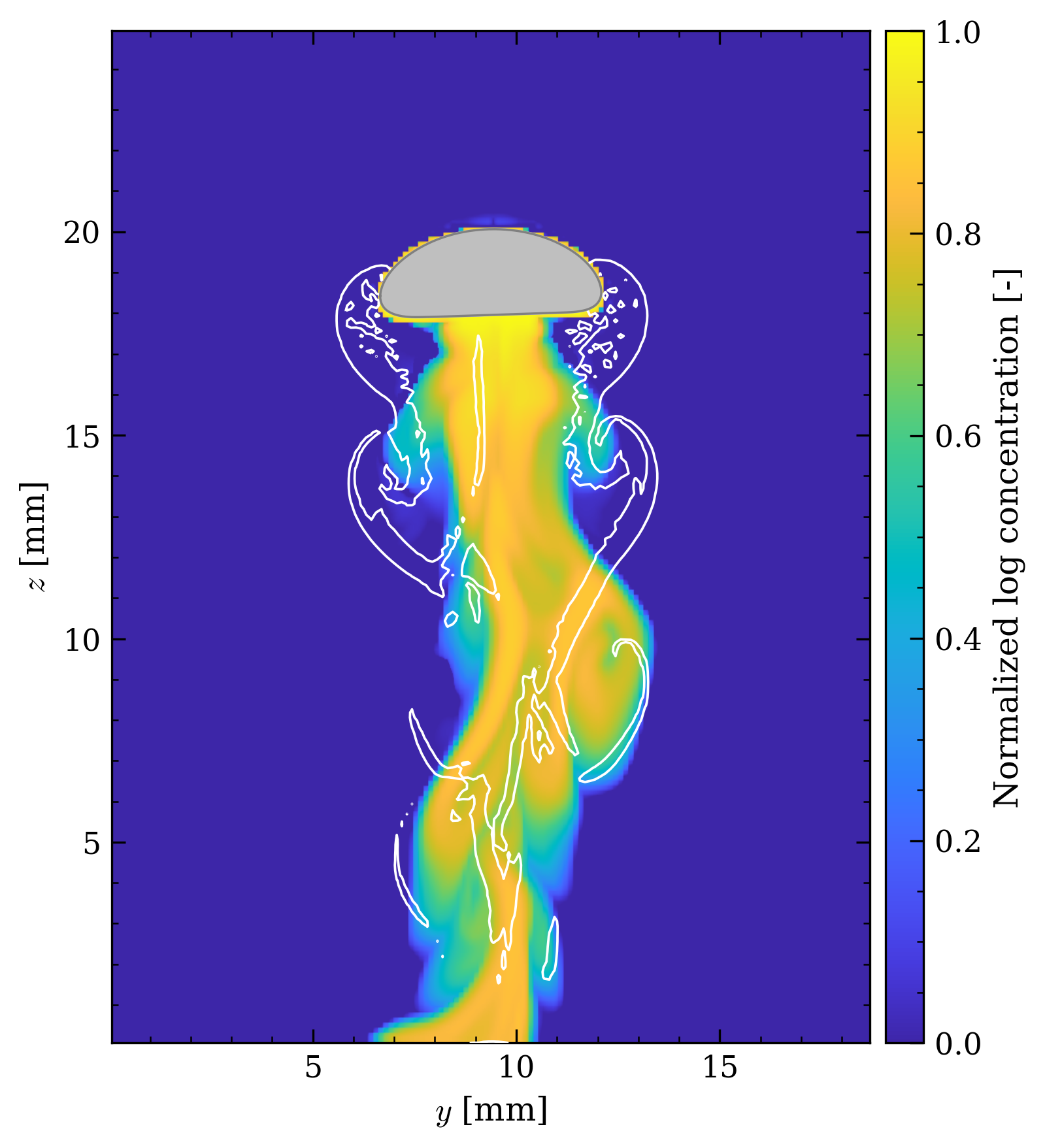}
        \caption{}
    \end{subfigure}
        \begin{subfigure}[]{0.32\textwidth}
        \centering
        \includegraphics[width=\textwidth]{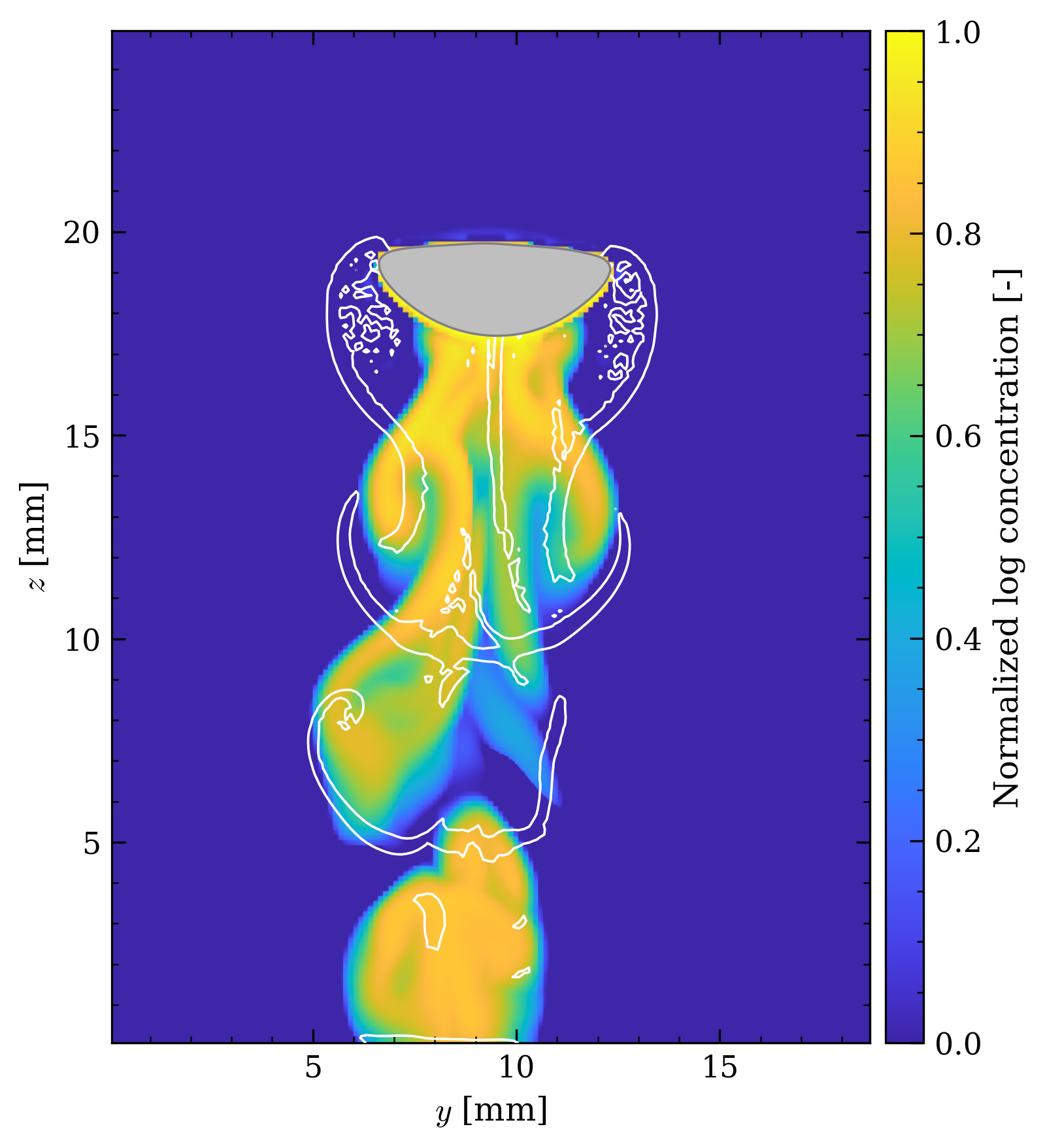}
        \caption{}
    \end{subfigure}
    \begin{subfigure}[]{0.32\linewidth}
        \centering
        \includegraphics[width= \textwidth]{figures/ftle_fwd_ridges_on_conc_yz.png}
        \caption{}
    \end{subfigure}
    \begin{subfigure}[]{0.32\linewidth}
        \centering
        \includegraphics[width= \textwidth]{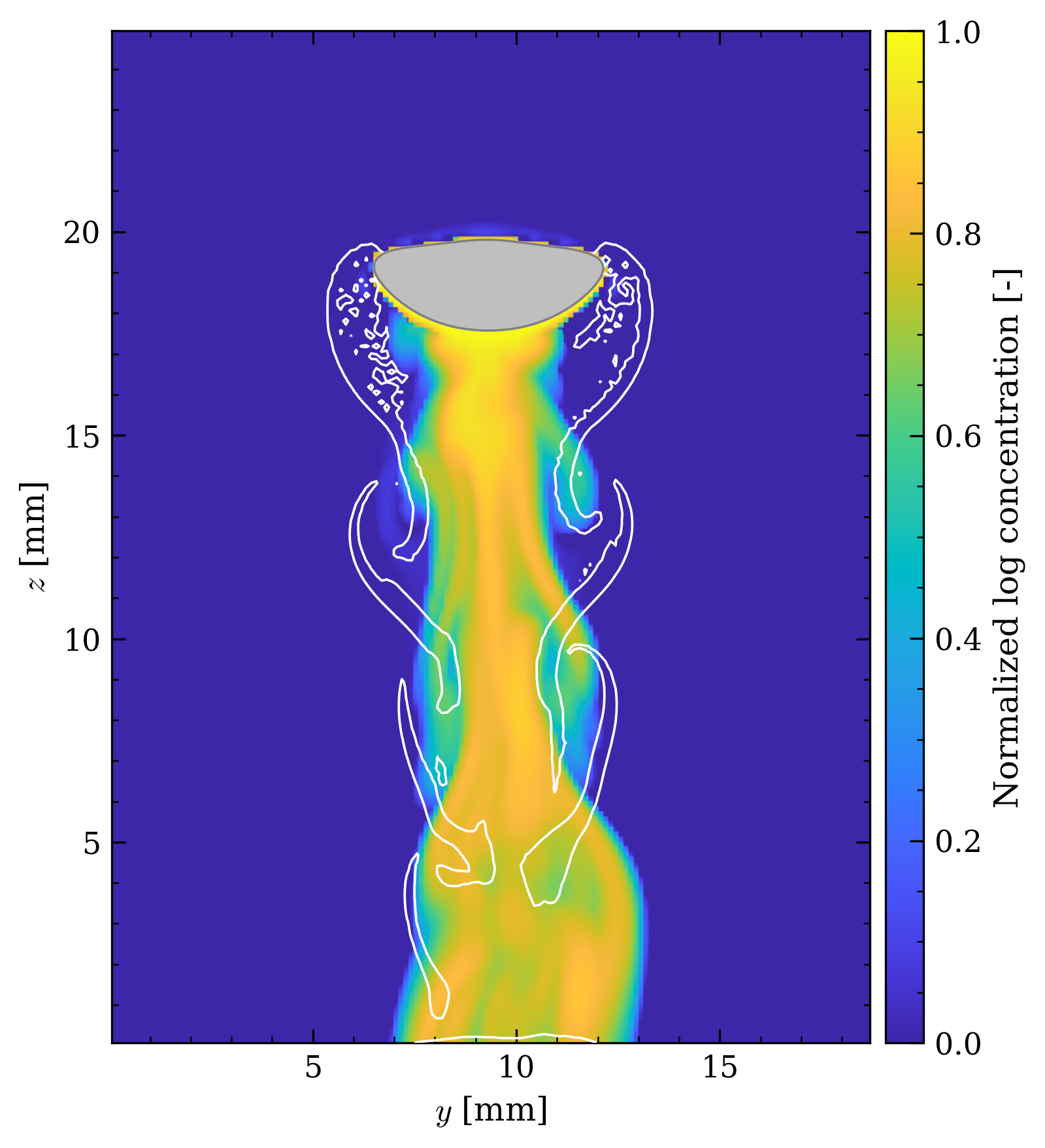}
        \caption{}
    \end{subfigure}
        \begin{subfigure}[]{0.32\textwidth}
        \centering
        \includegraphics[width=\textwidth]{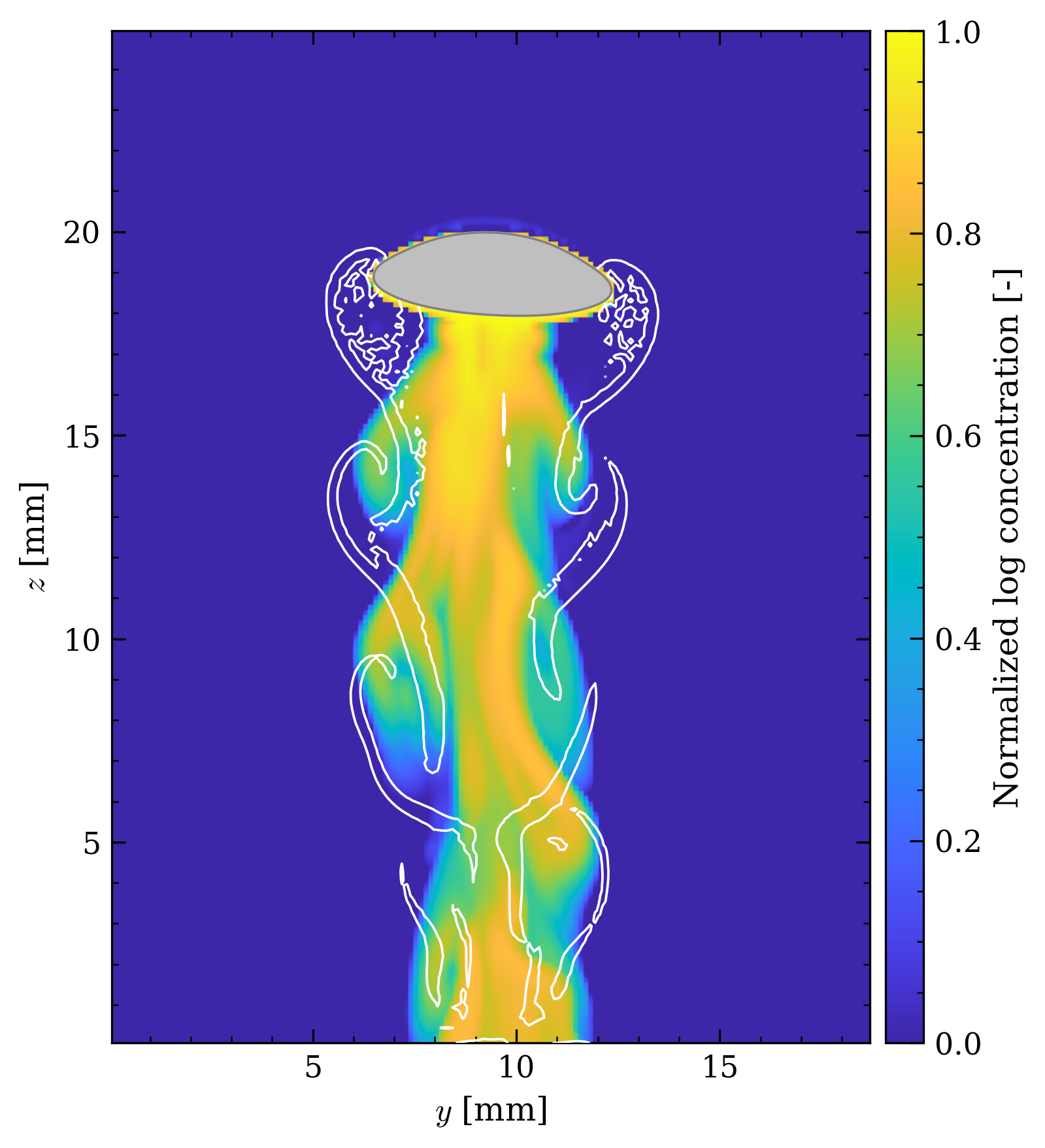}
        \caption{}
    \end{subfigure}
    \begin{subfigure}[]{0.32\linewidth}
        \centering
        \includegraphics[width= \textwidth]{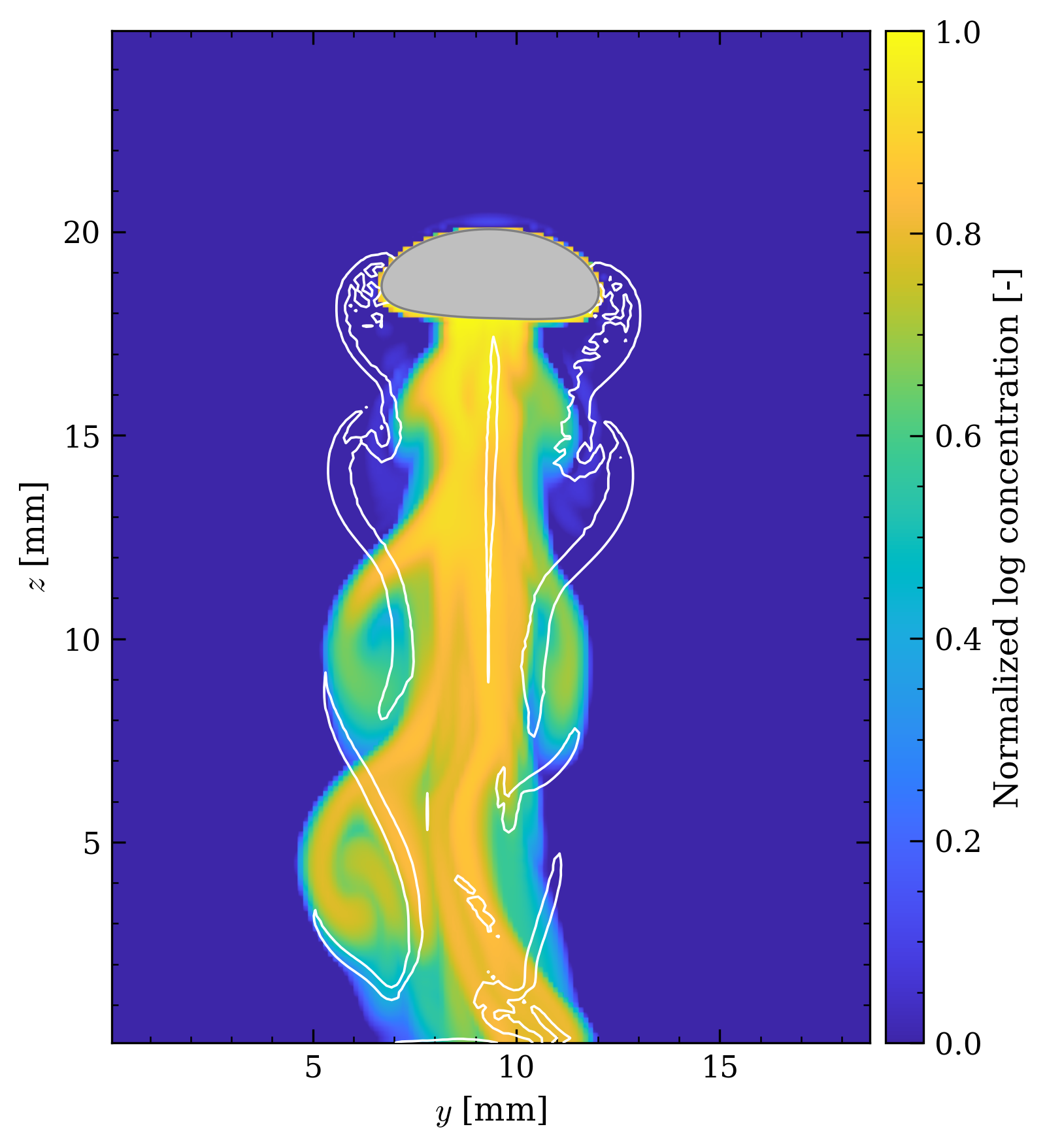}
        \caption{}
    \end{subfigure}
    \begin{subfigure}[]{0.32\linewidth}
        \centering
        \includegraphics[width= \textwidth]{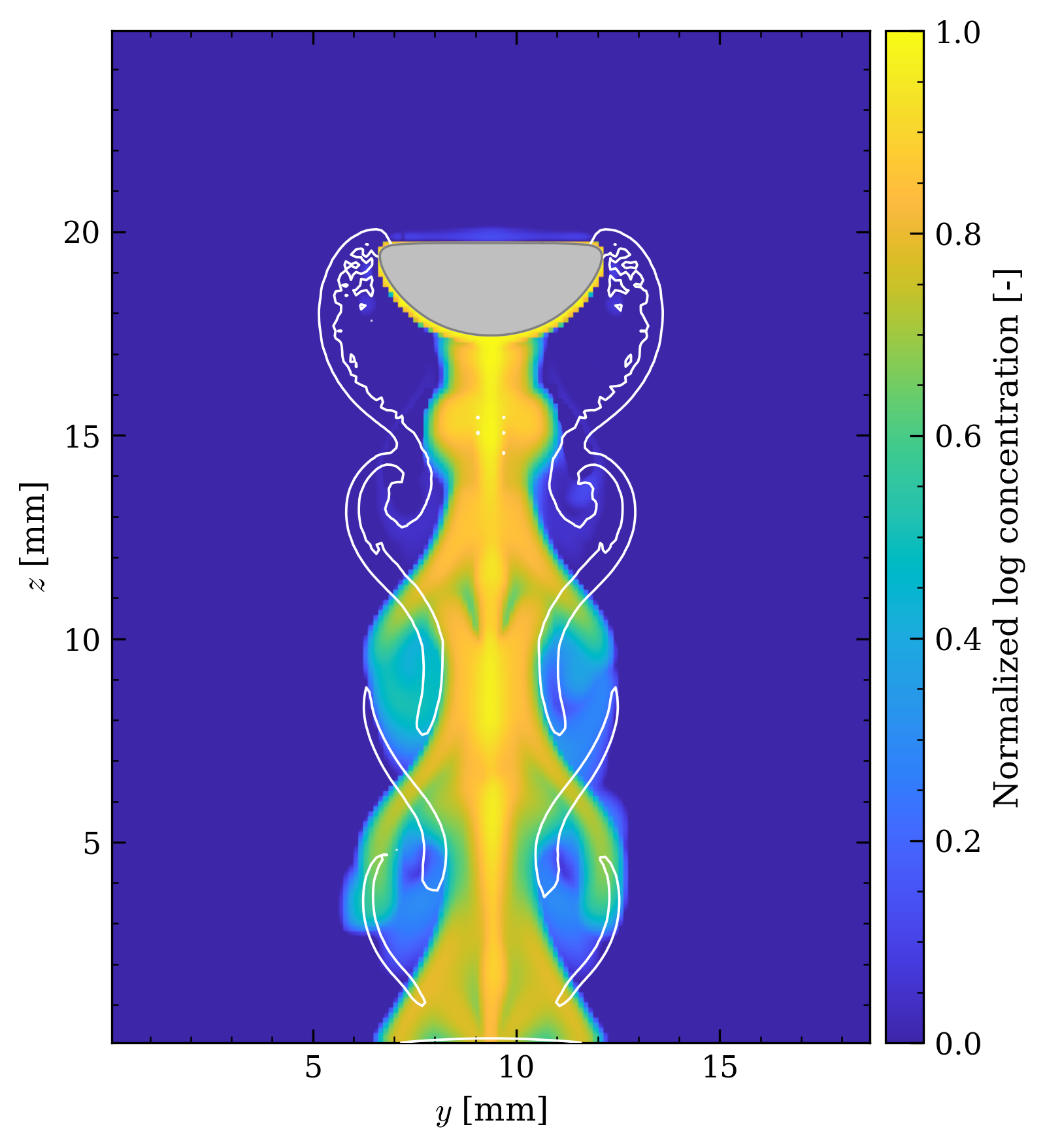}
        \caption{}
    \end{subfigure}
        \begin{subfigure}[]{0.32\textwidth}
        \centering
        \includegraphics[width=\textwidth]{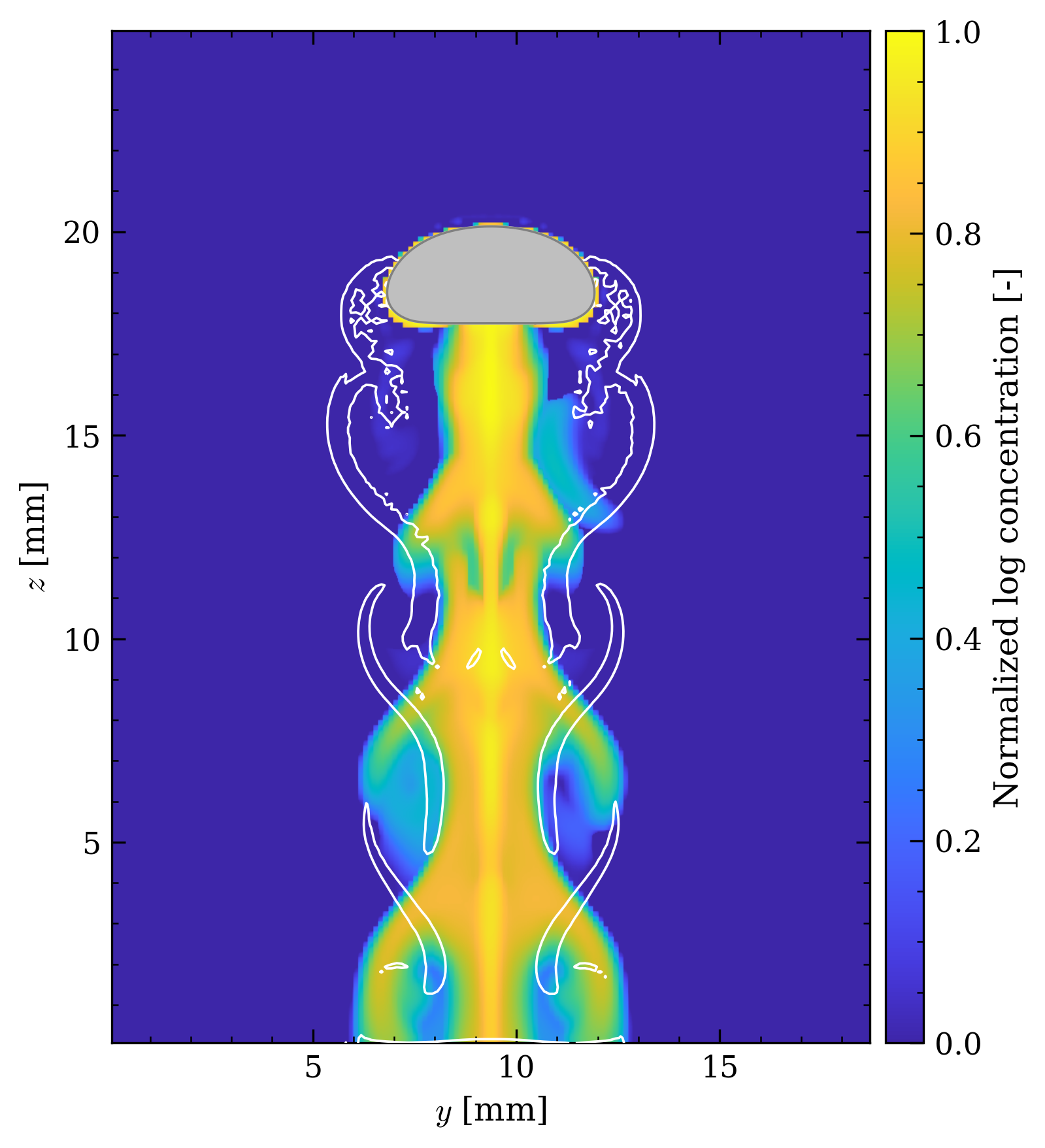}
        \caption{}
    \end{subfigure}
    \begin{subfigure}[]{0.32\linewidth}
        \centering
        \includegraphics[width= \textwidth]{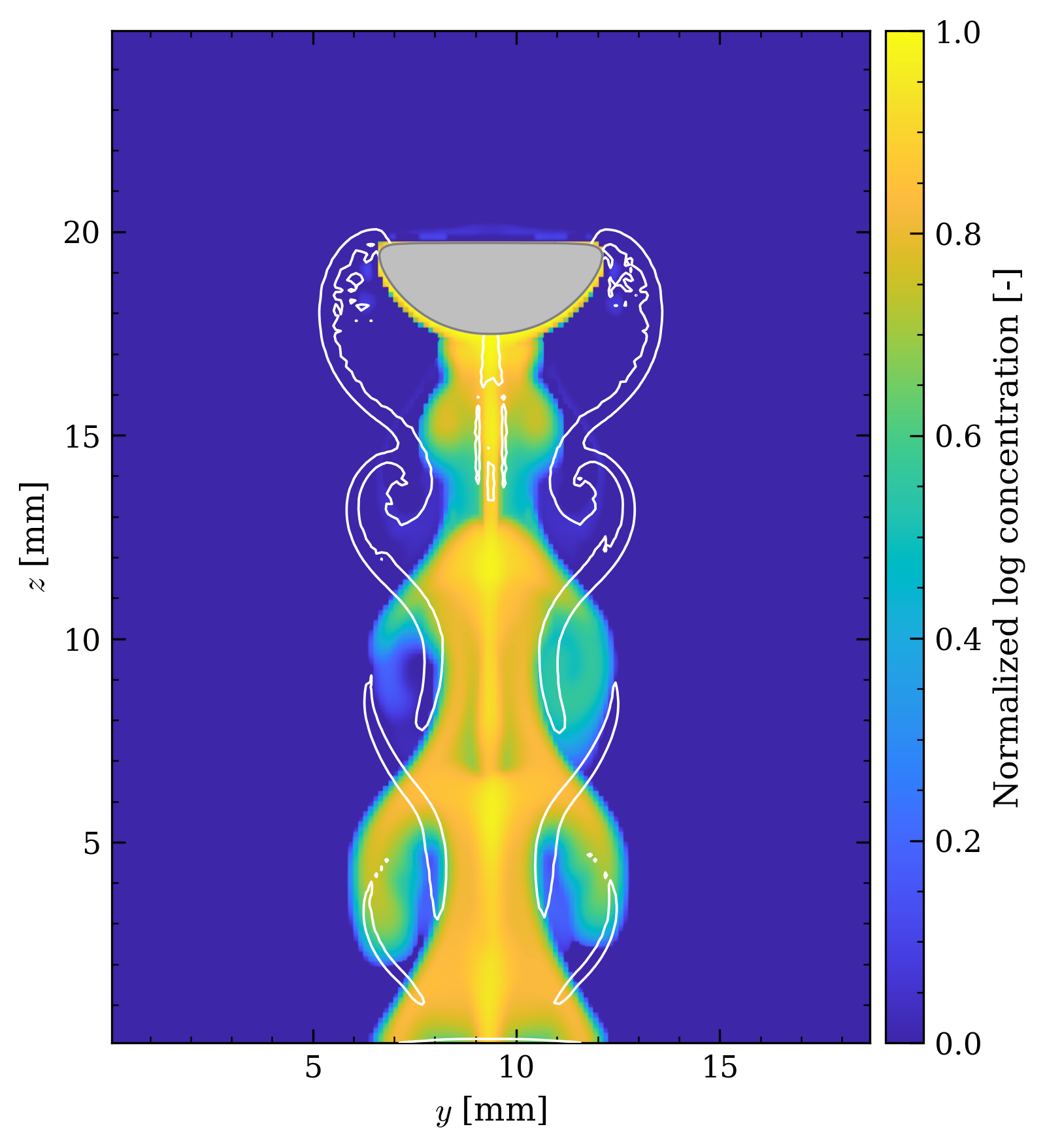}
        \caption{}
    \end{subfigure}
    \caption {The concentration profiles for case 5 (top row, $\kappa=\frac{\mu_g}{\mu_l}=10^{-1}$ and $\lambda=\frac{\rho_g}{\rho_l}=2\cdot10^{-1}$), case 6 (middle row, $\kappa=\frac{\mu_g}{\mu_l}=5\cdot10^{-1}$ and $\lambda=\frac{\rho_g}{\rho_l}=2\cdot10^{-1}$) and case 12 (bottom row, $\kappa=\frac{\mu_g}{\mu_l}=10^{-1}$ and $\lambda=\frac{\rho_g}{\rho_l}=2\cdot10^{-3}$). The lines in all figures indicate the FTLE ridges evaluated as the 5\% highest FTLE values. The different columns represent different time instances in the pseudo-steady-state.}
    \label{fig:fwd_conc_all}
\end{figure}

To determine the influence of the physical properties of the gas and liquid, the highest 5\% of forward FTLE values with the concentration fields of case 5, 6 and 12 are plotted in figure \ref{fig:fwd_conc_all}. Cases 6 and 12 are chosen because they have a significant decrease in the dynamic viscosity of the gas and a significant decrease in the density of the gas, respectively. Note that the backward FTLE ridges were also determined and show clear encapsulation of the high concentration regions as was shown in figure \ref{fig:case5_visualisation_bwd}. Therefore, the backward FTLE ridges are not shown here. 

Figure \ref{fig:fwd_conc_all} shows that in all cases the FTLE ridges align in the vertical direction, preventing mixing between the bulk and the bubble wake. In addition, both cases 6 and 12 have more clear vortical structures in the height of the domain. Combining this with the observation that the frequency of changes in the Reynolds number indicates that the higher oscillation frequency indicates not only a faster shedding of the vortices but also the creation of smaller vortices. 

In addition to the number of vortical structures, the structures in case 12 are more consistent throughout the domain compared to the other two cases. This indicates that the vortical structures in this case are more coherent and will sustain a longer period of time compared to the structures in the other cases. This difference could be caused by the slightly higher dynamic viscosity of the liquid in case 12 compared to cases 5 and 6, which would also result in a resistance to change the velocity profile.

In conclusion, the FTLE ridges show that clear convective transport barriers exist between the wake and the bulk of the liquid, which are more persistent when the dynamic viscosity of the liquid increases. In addition, several regions are enclosed by ridges of high FTLE values, which prevent convective mixing in the wake of the bubble. However, it should be noted that although convective transport will not occur through these FTLE ridges, diffusive transport can occur, especially when the concentration difference between both sides of the ridges is large. Therefore, FTLE analysis will not directly provide a measure for the concentration profile but will indicate the convective transport, which is the dominant transport mechanism.

\section{Conclusion}
In this work, 15 simulations were performed to study the effect of the wake structures on mass transfer over the gas-liquid interface and the concentration profile behind the wake. All simulations are performed for a wobbling bubble, indicated with an E\"otv\"os number of 2 and a Morton number of $10^{-11}$ ensuring that the bubble rise velocity indicated by the Reynolds number varied only by less than 3\% and thus eliminates the main effect of the bubble rise velocity on the mass transfer.

The results showed that the oscillations and thus the vortical structures created by the bubble are significantly influenced by the physical properties. This is shown not only in the oscillation frequency of the Reynolds number but also in the FTLE fields. The oscillations in the bubble rise velocity also lead to oscillations in the Sherwood number, effectively indicating that the oscillations in the velocity result in changes in the mass transfer. This can be explained by an increased surface renewal rate as a result of the change in shape and path of the bubble. 

Although the average Reynolds number is not affected, the average Sherwood number is affected by the physical properties of the gas and liquid, which can be attributed to small changes in the velocity profile in the vicinity of the gas-liquid interface. In addition, the amplitudes of the oscillations of the Reynolds number and Sherwood number are not correlated. This indicates that the local and global mass transfer over the gas-liquid interface are influenced by the dynamics of the bubble.

Finally, the FTLE analysis showed vertical ridges that prevent convective transport between the wake of the bubble and the bulk of the liquid; \textit{i.e.}, there is no mixing between the liquid in the wake and the bulk of the liquid. In addition, the FTLE ridges also show separation of small liquid volumes, which prevents mixing of the entire wake. Therefore, mass transfer at the gas-liquid interface cannot be regarded as mass transfer to the bulk of the liquid because a second mass transfer barrier occurs at the boundary of the wake. However, it should be noted that this is a purely convective transport barrier, and diffusive mass transfer can still take place between the wake and the bulk.

Although these different mechanisms can be identified, the application of these findings requires extension to different bubble sizes and liquids, bubble interactions, and the effect of turbulent  structures in the liquid. Only when these effects are included, closures can be obtained that allow for a better prediction of the mass transfer from a bubble to the bulk of the liquid.

The present study employs a stationary velocity field in the wake of the bubble, resulting in static FTLE fields. In reality, the motion and deformation of the bubble continuously modify the surrounding flow field, which may in turn affect the FTLE ridges and the associated transport barriers. Investigating the influence of a time-varying velocity field on these structures would therefore be a valuable direction for future research, which requires velocity fields over the integration period $\tau$.

\section*{Acknowledgements}
The authors thank Prof. Dr. Alexandra von Kameke for the valuable discussions regarding the mass transfer and velocity profiles of single bubbles.

\section{Appendix}
\begin{longtable}[!h]{lcccccccccccc}
    \caption{All simulations performed for the determination of the correlation.}
    \label{tab:allsim}\\
        \hline
        \hline
        case &$\mu_l$ &$\kappa=\frac{\mu_g}{\mu_l}$& $\lambda=\frac{\rho_g}{\rho_l}$& $\sigma$ (N/m)& $Re=\frac{\rho_l v d_{eq}}{\mu_l}$  & $f_{Re}$ (Hz)& $A_{Re}$&$Sh=\frac{\mu_l}{\rho_l D}$ & $f_{Sh}$ (Hz) &$A_{Sh}$ \\
         \hline
         \endfirsthead
         
         \hline
         \multicolumn{10}{c}{Continuation of table \ref{tab:allsim}}\\
        \hline
         case & $\mu_l$ (Pa s) $\kappa=\frac{\mu_g}{\mu_l}$& $\lambda=\frac{\rho_g}{\rho_l}$& $\sigma$ (N/m)& $Re=\frac{\rho_l v d_{eq}}{\mu_l}$  & $f_{Re}$ (Hz)& $A_{Re}$&$Sh=\frac{\mu_l}{\rho_l D}$ & $f_{Sh}$ (Hz) &$A_{Sh}$ \\
         \hline
         \endhead
         
         \hline
         \endfoot

         \hline
         \endlastfoot
         
         1 & $5.98\cdot10^{-4}$ & $2\cdot10^{-1}$ & $5\cdot10^{-1}$ & 0.0400 & 1293 & 22 & 256.6 & 1014 & 22 & 194.4 \\
         2 & $5.98\cdot10^{-4}$ &$2\cdot10^{-2}$ & $5\cdot10^{-1}$ & 0.0400 & 1296 & 22 & 190.3 & 1026 & 22 & 173.0 \\
         3 & $7.57\cdot10^{-4}$ &$5\cdot10^{-1}$ & $2\cdot10^{-1}$ & 0.0640 & 1303 & 32 & 252.2 & 1049 & 32 & 186.0 \\
         4 & $7.57\cdot10^{-4}$ &$2\cdot10^{-1}$ & $2\cdot10^{-1}$ & 0.0640 & 1305 & 32 & 290.3 & 1052 & 32 & 181.9 \\
         5 & $7.57\cdot10^{-4}$ &$1\cdot10^{-1}$ & $2\cdot10^{-1}$ & 0.0640 & 1292 & 32 & 300.5 & 1080 & 32 & 155.8 \\
         6 & $7.57\cdot10^{-4}$ &$5\cdot10^{-2}$ & $2\cdot10^{-1}$ & 0.0640 & 1311 & 36 & 173.1 & 1046 & 36 & 133.6 \\
         7 & $8.24\cdot10^{-4}$ &$2\cdot10^{-1}$ & $5\cdot10^{-2}$ & 0.0760 & 1292 & 38 & 355.2 & 1077 & 38 & 185.1 \\
         8 & $8.42\cdot10^{-4}$ &$2\cdot10^{-1}$ & $1\cdot10^{-2}$ & 0.0792 & 1257 & 40 & 451.3 & 1092 & 40 & 229.1 \\
         9 & $8.44\cdot10^{-4}$ &$2\cdot10^{-1}$ & $5\cdot10^{-3}$ & 0.0796 & 1267 & 41 & 463.9 & 1075 & 41 & 240.8 \\
         10 & $8.44\cdot10^{-4}$ &$2\cdot10^{-2}$ & $5\cdot10^{-3}$ & 0.0796 & 1252 & 41 & 463.1 & 1094 & 41 & 223.0 \\
         11 & $8.45\cdot10^{-4}$ &$5\cdot10^{-1}$ & $2\cdot10^{-3}$ & 0.0798 & 1264 & 41 & 431.0 & 1089 & 41 & 227.9 \\
         12 & $8.45\cdot10^{-4}$ &$1\cdot10^{-1}$ & $2\cdot10^{-3}$ & 0.0798 & 1254 & 37 & 429.8 & 1103 & 37 & 222.6 \\
         13 & $8.45\cdot10^{-4}$ &$5\cdot10^{-2}$ & $2\cdot10^{-3}$ & 0.0798 & 1253 & 38 & 436.6 & 1100 & 38 & 216.6 \\
         14 & $8.45\cdot10^{-4}$ &$2\cdot10^{-1}$ & $1\cdot10^{-3}$ & 0.0799 & 1257 & 37 & 421.2 & 1103 & 37 & 213.1 \\
         15 & $8.45\cdot10^{-4}$ &$2\cdot10^{-2}$ & $1\cdot10^{-3}$ & 0.0799 & 1255 & 41 & 425.0 & 1099 & 41 & 213.2 \\

\end{longtable}
\bibliographystyle{elsarticle-harv} 
\bibliography{bibliography}





\end{document}